\documentclass[12pt,letterpaper]{article}
\usepackage{amsmath}

\usepackage{amsfonts}
\usepackage{graphicx}
\usepackage{float}

\numberwithin{equation}{section}
 \renewcommand{\thefootnote}{\arabic{footnote}}
\newpage

\begin{document}
\begin{titlepage}
\hfill MCTP-14-41\\

\begin{center}
{\Large \bf Homothety and Quasi Self Similarity}\\
\vskip .7cm

{\Large \bf in Asymptotically Flat and AdS Spacetimes}
\end{center}

\vskip .7 cm

\vskip 1 cm
\begin{center}
{\large  Wenli Zhao}
\end{center}

\vskip .4cm \centerline{\it Michigan Center for Theoretical
Physics}
\centerline{ \it Randall Laboratory of Physics, The University of
Michigan}
\centerline{\it Ann Arbor, MI 48109-1120}

\vskip 1 cm

\begin{abstract}
For a spherically symmetric self-gravitating scalar field we study self similar and quasi-self similar solutions in asymptotically flat and AdS spacetimes in various dimensions. Our main approach relies on reducing the Einstein-Klein-Gordon system of equations under the self similarity assumption to a dynamical system. Applying standard techniques in the analysis of dynamical system we study the corresponding phase diagrams, gaining some insight into the short and long-term evolution of the system. In the asymptotically flat case the fixed points of the dynamical system correspond to conformally flat spacetimes. We observed a new transition from a fixed point to a black hole and we compute the corresponding exponent analytically. In the case of AdS spacetime, we observed the quasi self similar scalar field scattering to infinity after interfering near the center corresponding to the dynamical system with no fixed points.
\end{abstract}

\end{titlepage}

\setcounter{page}{1} \renewcommand{\thefootnote}{\arabic{footnote}}
\setcounter{footnote}{0}
\def \N{{\cal N}}
\def \ov {\over}

\section{Introduction}

The study of self-similar solutions in the context of Einstein's gravity has a long and fruitful history (see \cite{Carr:1998at}  for a review and  references). There are various motivations to study self-similar solutions.Christodoulu used self similarity to produce examples of scalar field solutions containing naked singularities \cite{Christodoulou1994}, and Choptuik in his remarkable discovery pointed to the possibility of certain self-similarity in the threshold of black hole formation \cite{Choptuik:1992jv}. 

Similar critical behavior has also been established numerically in many systems, some of which contain explicit scales \cite{Gundlach:2002sx}\cite{Gundlach:1999cu}. It is clear that the self-similarity present during critical collapse is discrete. Nevertheless it has inspired work on continuous self-similarity. 

In contrast to the fruitful results of self similar collapse in asymtotically flat spacetime, only a few results have been obtained for self similar scalar field solution in Anti-de-Sitter(AdS) spacetime. An absence of a self similar solution in $AdS$ is not surprising, since the presence of the $AdS$ radius sets the scale of the system. But there still can be approximate forms of self smilarity in this context.  Clement Gerard and Fabbri Alessandro observed the quasi self similar solutions in $AdS_3$\cite{Clement:2001ak} and R Baier, S A Stricker and O Taanila provided a concrete analytical example in their recent paper\cite{Baier:2014}. However, the generalization of quasi self similar scalar field solutions to higher dimensions is not clear. 

Our main motivation is to shed some light into the nature of the critical behavior during gravitational collapse in higher dimensional AdS spacetime. Different from the analytical approaches employed in $AdS_3$, our approaches lie on standard dynamical system techniques, which have been introduced by Brady\cite{PhysRevD.51.4168} and extensively used for case of asymtotically flat space. We exhibited the possiblity of obtaining semi analytical understanding even in the case of higher dimensional AdS. 

The paper is organized as follows. In section \ref{Sec:Flat} we discuss self-similarity in asymptotically flat spacetimes. We follow the standard steps in turning the problem into a dynamical system, but we observe a new transition from a fixed point to a black  hole and we compute the critical exponent in this case. In section \ref{Sec:AdS} we show that asymptotically AdS spacetimes regardless of the dimensions do not admit self-similar solutions but admit a possibility for quasi self-similarity that allows for an analysis of the gravitational system in the language of dynamical systems. We find no fixed point for such a system and discuss its implications. We finish with consclusions in section \ref{Sec:Conclusions}. We relegate some technical details to the appendices.

\section{Homothetic gravitational collapse of a spherically symmetric scalar field in $n$-dimensional flat space-time}\label{Sec:Flat}

In this section we follow the standard steps (such as those in \cite{PhysRevD.51.4168}) to turn the higher dimensional  Einstein Klein Gordon system into a dynamical system in the presence of a parameter $k$ that stands for the additional freedom given by the scalar field. Our system agrees with that of Brady\cite{PhysRevD.51.4168} when $n=4$ and with that of Soda \cite{Soda1996271} when $k=0$. We provide a breif discussion on the classification of the solutions and exhibit a new transition from fixed point to blackholes in solutions of Class One. Concrete phase diagrams are provided for $n=5$ and $n=6$.

\subsection{The Einstein Klein Gordon system as a dynamical system}
For the spherically symmetric setup whereby the metric takes the form
\begin{equation}\label{metric}
ds^2= g_{ab}dx^a dx^b+r^2d\Omega^2,
\end{equation}
where $a,b$ run over the coordinate on the quotient manifold, $\Omega$ denotes the solid angle and $r$ the area radius function in the sense of \cite{CPA:CPA3160460803}. For an $n$-dimensional spacetime, the evolution of the system is determined by the following equations on the quotient manifold with boundary, where the boundary is given by a set of time-like geodesic.  
\begin{equation}\label{Einsteineq}
(n-2)r\nabla_a\nabla_b r = \frac{(n-2)(n-3)}{2}g_{ab}(1-\partial^c r\partial_c r)-r^2T_{ab},
\end{equation}
\begin{equation}\label{Energymomentumeq}
T_{ab}=\partial_a\phi\partial_b\phi-\frac{1}{2}g_{ab}\partial^c\phi\partial_c\phi,
\end{equation}
where $\nabla$ stands for covariant derivative, and $T_{ab}$ the energy momentum tensor. Using $r$ and a null coordinate $u$, the metric on the quotient can be written as
\begin{equation}\label{Bondimetric}
g_{ab}dx^adx^b= -g\bar{g}du^2-2gdudr,
\end{equation}
where $g$ and $\bar{g}$ are functions of $u$ and $r$ only. 

Observe that the system is invariant under the following transformation
\begin{equation}\label{transformation}
g_{ab}\mapsto c^2 g_{ab}, \qquad r\mapsto c\,\,r, \qquad \phi \mapsto \phi + k \log c.
\end{equation}
Therefore, it is natural to seek a solution that is scale invariant of type $ k$ for any $k\in \mathbb{R}$ in the sense of \cite{1994}. That is, there exists a family of diffeomorphisms $\{f_a \}$ such that $f_a^{*}(g_{uv})=a^2g_{uv}, f_a^{*}(r)=ar$ and $f_a^{*}(\phi)=\phi+k\log{a}$. 
In \cite{1994} it is proven that in a  four-dimensional spacetime, the metric of quotient manifold in such a solution depends only on $\frac{r}{u}$. We can easily extended the proof to our case since we only deal with quantities on the quotient manifold. Therefore the system is invariant under the rescaling $r\mapsto \lambda r, u \mapsto \lambda u$ and in this sense it is self similar.

Using the Ansatz (\ref{Bondimetric}), the Einstein equations  are:
\begin{equation}\label{flatfieldeq1}
(n-2)\frac{\partial_r g}{rg}=(\partial_r \phi)^2,
\end{equation}
\begin{equation}\label{flatfieldeq2}
\partial_u(\frac{\bar{g}}{g})=\frac{2r}{(n-2)g}((\partial_u \phi)^2-\bar{g}\partial_u\phi\partial_r\phi),
\end{equation}
\begin{equation}\label{flatfieldeq3}
\partial_r(r^{n-3}\bar{g})=(n-3)gr^{n-4},
\end{equation}
where $n$ is the dimension of the spacetime and $\phi$ is the scalar field. The Klein-Gordon equation becomes
\begin{equation}\label{flatfieldeq4}
\partial_r(r^{n-2}\bar{g}\partial_r \phi)=2r^{n-2}\partial_u\partial_r\phi+(n-2)r^{n-3}\partial_u \phi.
\end{equation}

The local mass, as \cite{Chatterjee} suggests, is directly related to the angular component of Riemann tensor under spherical symmetry. In our case, the local mass reduces to
\begin{equation}\label{localmass0}
m=\frac{r^{n-3}}{2}(1-\frac{\bar{g}}{g}).
\end{equation}
It agrees with the ADM and Bondi mass as $r\to \infty$ and $t\to \infty$. The positivity of the local mass requires $\frac{\bar{g}}{g}\leq 1$.

Take the Ansatz $\phi(\frac{r}{u})=\int_{0}^{\frac{r}{u}}\frac{\gamma(s)}{s} ds -k\ln{u}$ and let $x= \log(\frac{r}{|u|}),y=\frac{\bar{g}}{g}, z=\frac{r}{u\bar{g}} $.There are two patches of the solution: $u>0$ and $u<0$, but they are related by a reflection therefore without losing of generality we will only discuss $u<0$.

Equations (\ref{flatfieldeq1}-\ref{flatfieldeq4}) become an autonomous dynamic system as follows:
\begin{equation}\label{dynamiceq1}
\dot{z}=z(n-2-\frac{n-3}{y}),
\end{equation}
\begin{equation}\label{dynamiceq2}
\dot{y}=n-3-(\frac{\gamma^2}{n-2}+n-3)y,
\end{equation}
\begin{equation}\label{dynamiceq3}
\dot{\gamma}=\frac{(n-2)kz-(\frac{n-3}{y}-(n-2)z)\gamma}{1-2z}.
\end{equation}
One can show (\ref{flatfieldeq4}) is implied by equations (\ref{dynamiceq1}-\ref{dynamiceq3}). Also one notices that the system is intrinsically two dimensional since $\gamma$ can be related to $z$ and $y$ by the following algebraic relation.
\begin{equation}\label{relation}
\gamma = -k \pm \sqrt{\frac{k^2+(n-2)(n-3)(1-y^{-1})}{1-2z}}.
\end{equation}
Equation (\ref{relation}) requires either $z\leq\frac{1}{2}$, $y\geq\frac{(n-2)(n-3)}{k^2+(n-2)(n-3)}$ or $z\geq\frac{1}{2}$, $y\leq\frac{(n-2)(n-3)}{k^2+(n-2)(n-3)}$. By (\ref{localmass0}) $y\in (0,1)$. Thus the physical regions for the autonomous system are $z\geq\frac{1}{2}, 0\leq y\leq\frac{(n-2)(n-3)}{k^2+(n-2)(n-3)}$ and $z\leq\frac{1}{2},\frac{(n-2)(n-3)}{k^2+(n-2)(n-3)}\leq y\leq 1$

Note that $y\to 0$ implies $y=\frac{\bar{g}}{g}=g^{uv}\partial_u r\partial_v r\to 0$ which signals an apparent horizon.

The stress energy tensor takes the form

\begin{equation}\label{energy stress tensor at the origin}
T_{uv}\sim \frac{\gamma^2}{r^2}.
\end{equation}
To avoid a singularity at the origin $r\to 0$, we need to impose a condition on the function $\gamma$:
\begin{equation}\label{gammainitial}
\gamma(0) =0 .
\end{equation}

A similar dynamical system for gravitational collapse in higher dimensions limited to $k=0$ was obtained in \cite{Soda1996271}  and was thoroughly discussed.We incoporate the additional parameter $k$ and we will show in the next section that it is relevant to critical collapse.




\subsection{Classification of solutions and discussions}

The non-trivial fixed points of  the autonumous system given by Eqs. (\ref{dynamiceq1}-\ref{dynamiceq3}) are given as follows in terms of $(z,y,\gamma)$
\begin{equation}\label{fixedpoints}
(\frac{\sqrt{n-2}}{k+\sqrt{n-2}},\frac{n-3}{n-2},\sqrt{n-2}) \qquad \text{  or  } \qquad (-\frac{\sqrt{n-2}}{k-\sqrt{n-2}},\frac{n-3}{n-2},-\sqrt{n-2})
\end{equation}

The fixed point represents conformally flat spacetimes and can be solved exactly, as we show in appendix (\ref{App:ConformalMetric}), and in appendix  \ref{App:Stability} we present some details of the fixed point analysis of the corresponding system.

\begin{figure}[htp]
\begin{center}
\includegraphics[width=2.5in,height=2.5in]{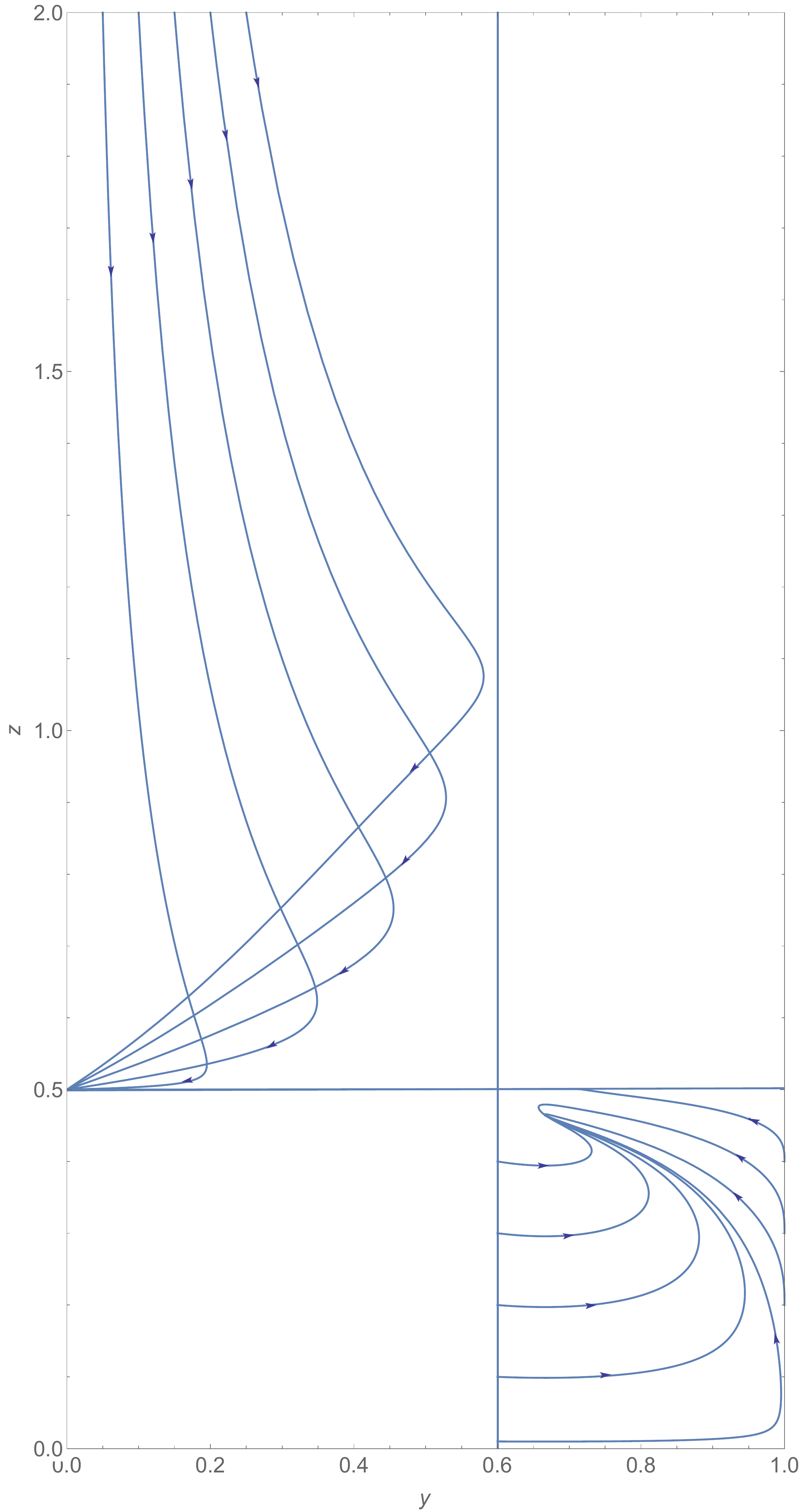}
\caption{\label{PhasePortrait1} Phase diagram for solutions of Class One. Here we take $k=2$, $n=5$. In this class there are two stable fixed points, one of which is shown on the diagram. The apparent horizon is given by the condition $y\to 0$.  On the one parameter initial data close to $y=1$ we observe a transition between non-black hole to black hole, which gives rise to the critical phenomena.}
\end{center}
\end{figure}

\begin{figure}[htp]
\begin{center}
\includegraphics[width=2.5in,height=2.5in]{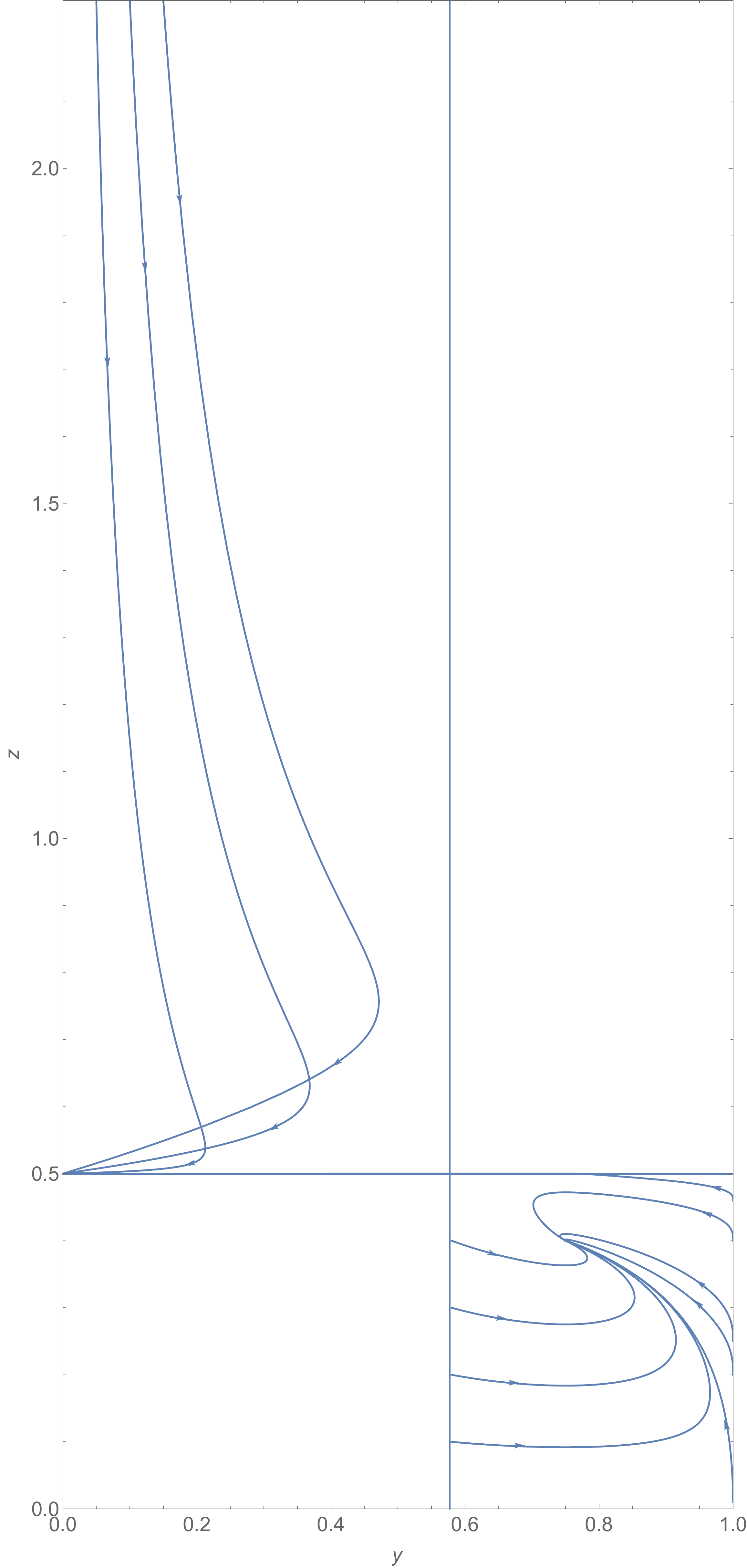}
\caption{\label{PhasePortrait3} Phase diagram for solutions of Class One. Here we take $k=3$, $n=6$. It is evident that $n=6$ case is similar to the $n=5$ case from our discussion above. Note that we also observed the transition from non-black hole to black hole for a one parameter data close to $y=1$. }
\end{center}
\end{figure}

Corresponding to the dynamical behaviors of the fixed points (\ref{fixedpoints}), there are two classes of solutions. For a given dimension, the first class is given by $k^2\geq n-2$. In this class, (\ref{fixedpoints}) are both stable. For a one-parameter data that is close to $y=1$, we observe a transition from non-black hole to black hole. In Fig.(\ref{PhasePortrait1}) and Fig.(\ref{PhasePortrait3}) we produce concrete phase diagrams for $n=5$ and $n=6$. 

\begin{figure}[htp]
\begin{center}
\includegraphics[width=2.5in,height=2.5in]{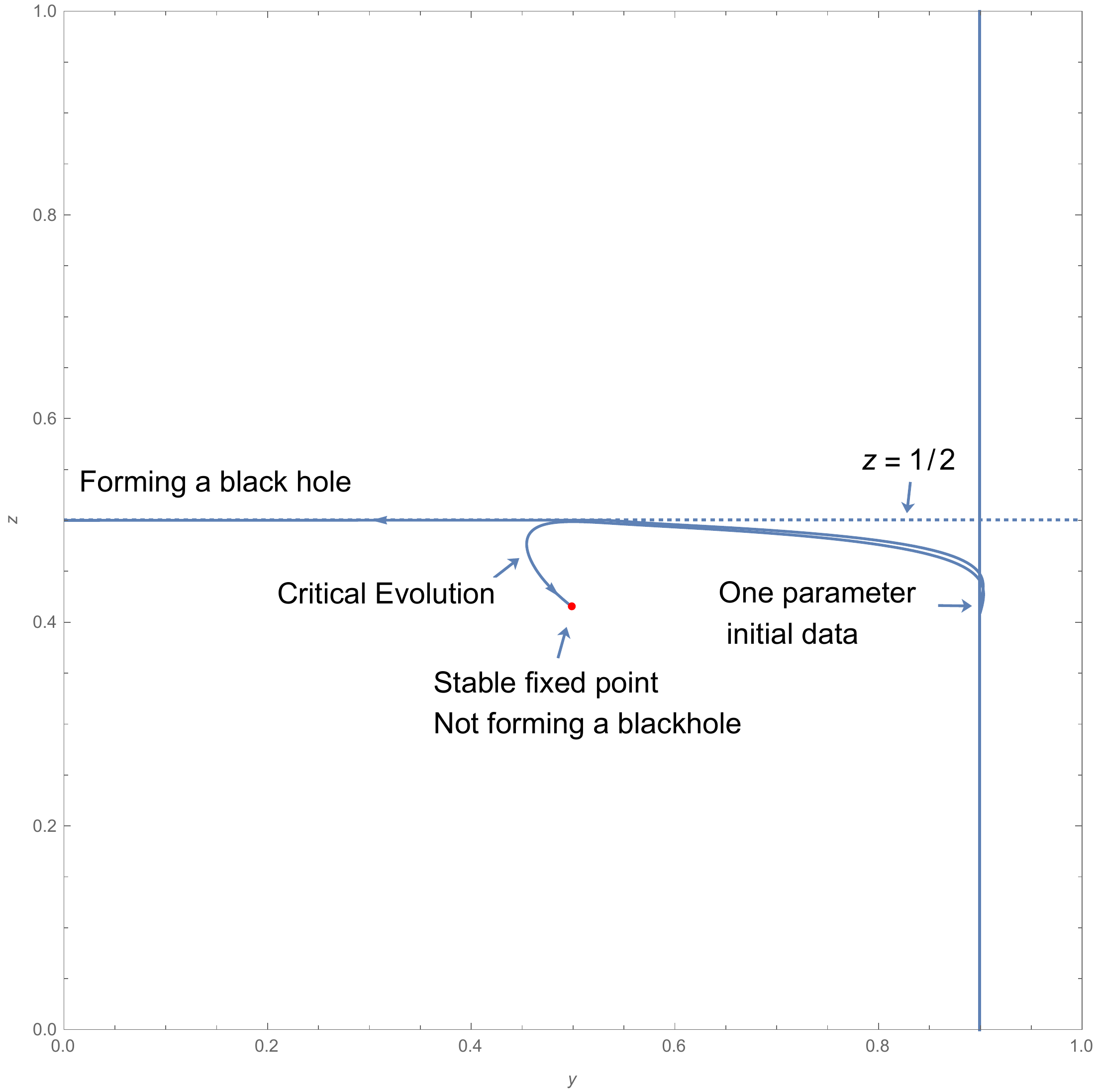}
\caption{\label{Cartoons}A qualitative picture of the transition between a fixed point and a black hole in the solutions of Class One. }
\end{center}
\end{figure}

Notice that in the first class, there is a transition from the fixed point to black holes for one parameter data near $y=1$, as the numerical evidence suggests. To make it clear, we reproduce the qualitative picture for this new transition in Fig.(\ref{Cartoons}). 

 Now we investigate this cirtical phenomnon analytically and compute the critical exponent, following the treatment givne by \cite{Soda1996271}.We expand asymtotically around the point $z=\frac{1}{2}, y=\frac{n-3}{n-2}$, since for each evolution $\dot{z}=0$ as $y=\frac{n-3}{n-2}$, which signals a turning point, and for initial $y>\frac{n-3}{n-2}$, $\dot{z}>0$. The critical behavior occurs when the evolution is near $z=\frac{1}{2}$ enough. Thus the critical point should be such that $z=\frac{1}{2}$ and $\dot{z}=0$, that is $z=\frac{1}{2}$ and $y=\frac{n-3}{n-2}$. Let $z=\frac{1}{2}(1-\delta)$ and $y=\frac{n-3}{n-2}+\epsilon$ and Eqs. (\ref{dynamiceq1}-\ref{dynamiceq3}) are turned to, approximated to the leading order,
\begin{equation}\label{asymtotical perturbation eq1}
\dot{\delta}=-(n-2)\epsilon
\end{equation}
\begin{equation}\label{asymtotical perturbation eq2}
\dot{\epsilon}=\frac{n-3}{n-2}-\frac{k^2(n-3)}{(n-2)^2}+\frac{4k^2(n-3)}{(n-2)^2}\delta-(\frac{5k^2}{n-2}+n-3)\epsilon
\end{equation}
Thus the eigen values are
\begin{equation}
\lambda_{1}=-(n-2), \lambda_2={\frac{4k^2(n-3)}{(n-2)^2}}
\end{equation}
Now we compute the critical exponent. Suppose $\lambda$ is the eigen value for the relevant mode.
The apparent horizon $r_{A.H}$ is of  the order
\begin{equation}\label{rAH}
r_{A.H} \sim (p-p*)^{\frac{1}{\lambda}},
\end{equation}

where $\lambda$ is the eigenvalue of the Jacobian of the autonomous system at the point where we expand the system, $p$ is the parameter for the one-parameter regular initial data and $p*$ is the critical threshold for the critical formation of a black hole. The black hole mass, subsequently,   has order
\begin{equation}\label{blackholemass}
m \sim r^{n-3} \sim (p-p*)^{\frac{n-3}{\lambda}}.
\end{equation}

Therefore, the exponent for a self similar type $k$ solution is given by
\begin{equation}\label{exponent}
\gamma=\frac{(n-2)^2}{4k^2}
\end{equation}

Note that this exponent is well-defined for $k \neq 0$. To our knowledge, we are thus presenting a new evolution with critical exponent different from those present in the literature\cite{Soda1996271}\cite{PhysRevD.51.4168}

\begin{figure}[h]
\begin{center}
\includegraphics[width=2.5in,height=2.5in]{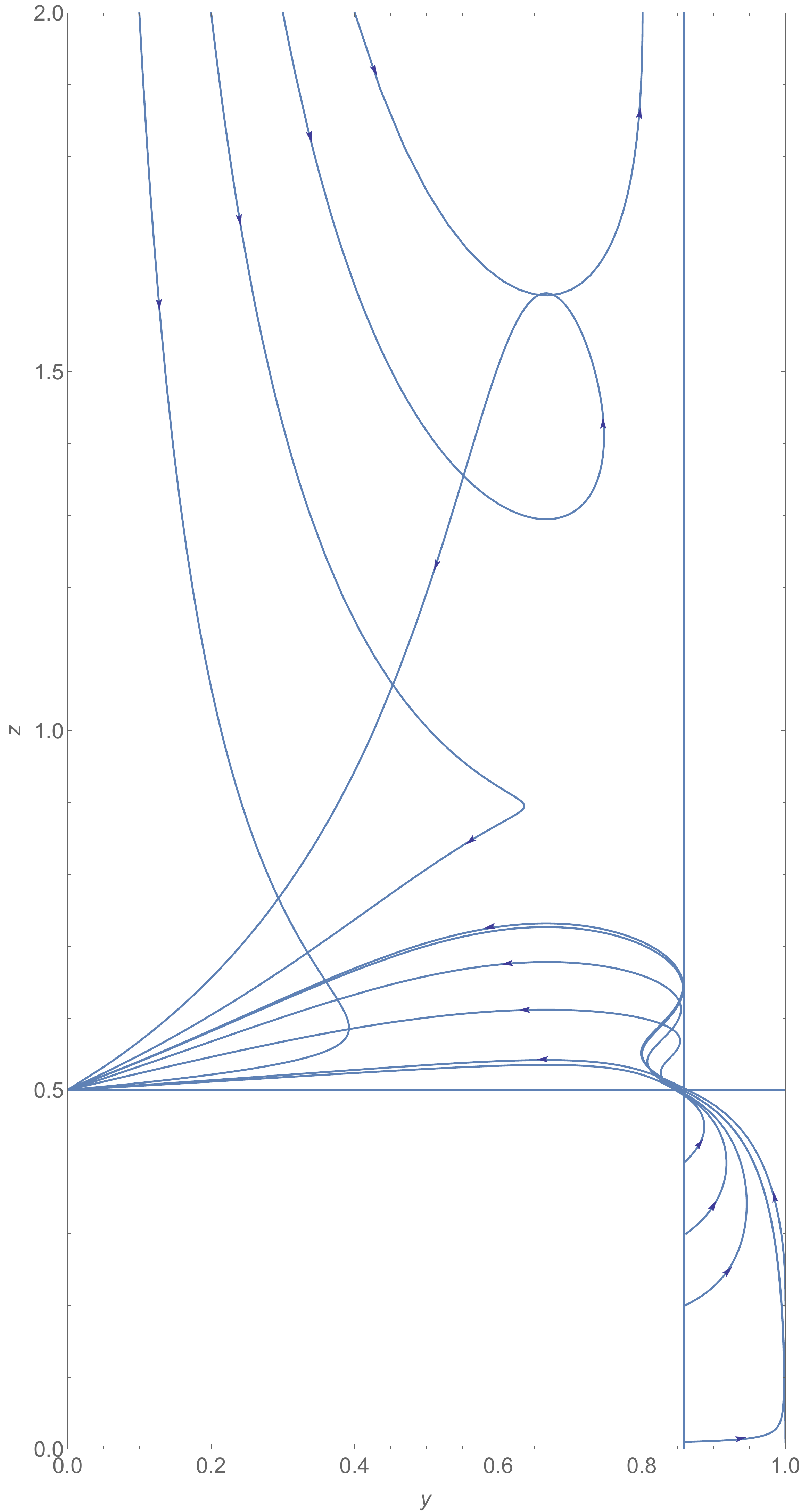}
\caption{\label{PhasePortrait2} Phase diagram for solutions of Class Two. Here we take $k=1$ and $n=5$. Note that in this class we can observe the transition between black hole and naked singularity ( $z\to \infty$).  }
\end{center}
\end{figure}

\begin{figure}[H]
\centering
\includegraphics[width=2.3in,height=2.5in]{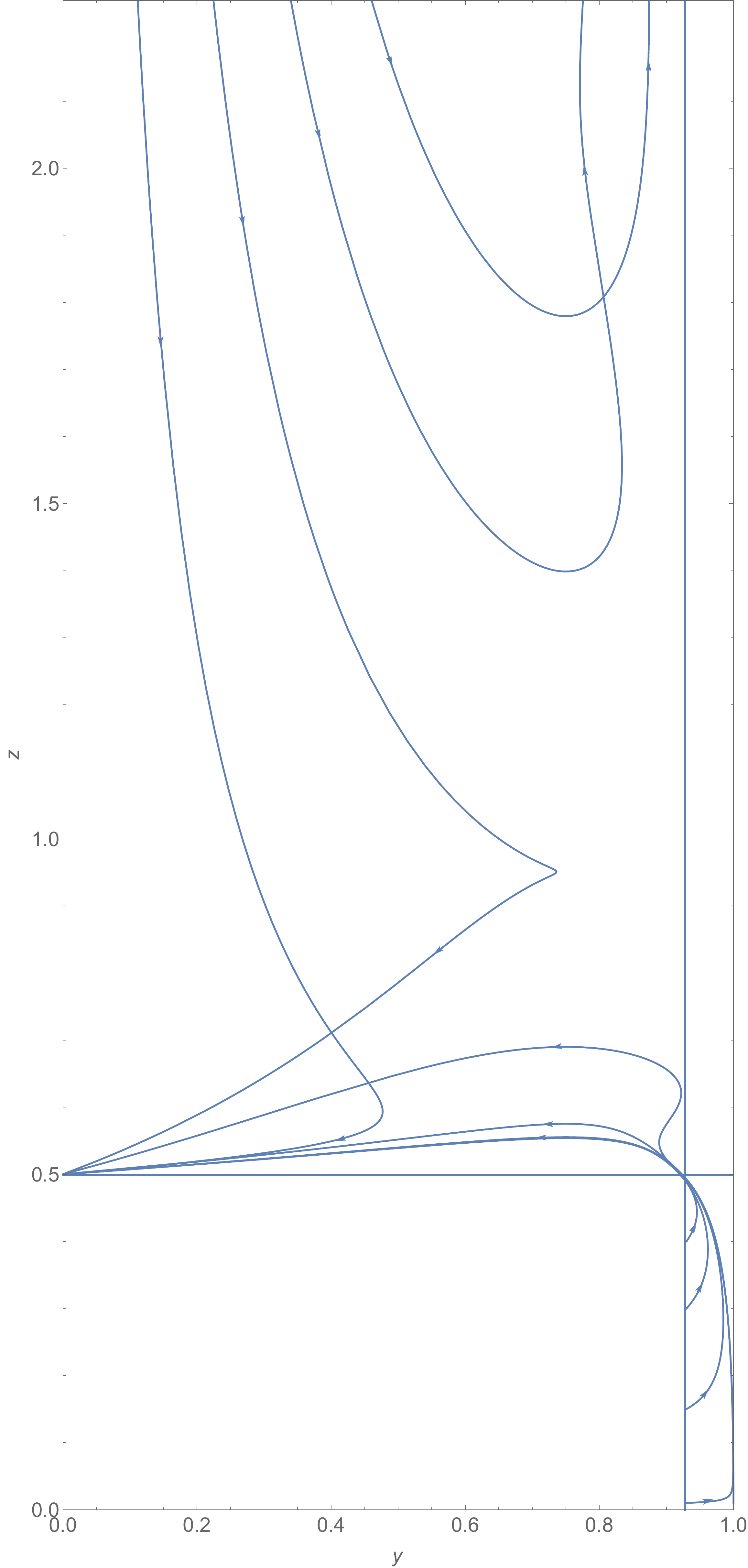}
\caption{\label{PhasePortrait4} Phase diagram for solutions of Class Two. Here we take $k=1$ and $n=6$.  It exhibits similar features to $n=5$ case. We can still observe the transition between black hole and naked singularity. }
\end{figure}

For completeness, we also incoporate the second class of solution which is given by $k^2< n-2$.  The fixed points (\ref{fixedpoints}) are both unstable in this class. In this class we can observe a transition from the formation of a black hole to that of a naked singularity regardless of the dimension of the spacetime. The figures Fig.(\ref{PhasePortrait2}) and Fig.(\ref{PhasePortrait4}) are the phase diagrams for $n=5$ and $n=6$ cases. 

\section{Quasi Self-similar gravitational collapse in $AdS_n$}\label{Sec:AdS}
In the previous section we discussed the new transition from a fixed point to a black hole in the case of self similar scalar field solutions in the asymtotically flat space, within the frame work of dynamical systems. It is natural to generalize that approach to higher dimensional $AdS$ and obtain a semi analytical description. Although several numerical works have observed self similar behavior in $AdS$ such as \cite{Bizon:2011gg} and \cite{deOliveira:2012dt}
, the $AdS$ radius provides the system with a fixed scale and thus does not allow for self similar solutions. The behaviors observed are actually in the regions where the $AdS$ radius is negligible. 

But one can, instead of asking for a self similar solution, ask for solutions that contain a self similar part with respect to some specific Ansatz. Those solutions are called quasi self similar solution and their presence, under suitable Ansatz, can still turn the PDE's into ODE's.  Several quasi self similar solutions have been found in $AdS_3$ by \cite{Clement:2001ak} and one particular case has been discussed by \cite{Baier:2014}. 

In this section, inspired by \cite{Clement:2001ak} and \cite{Baier:2014}, we first motivate the choice of Ansatz through the light-cone symmetry that the system possesses. Instead of pursuing analytical solutions, we turn the quasi self similar Einstein Klein Gordon system in higher dimensional $AdS$ into a dynamical system and present a semi-analytical understanding of the interactions of the scalar field that scatters to infinity in a finite amount of time.

\subsection{The breaking of homothetic symmetry in AdS}
We now consider the Einstein-Klein-Gordon system in $ AdS_n$. The equations on the quotient manifold are
\begin{equation}\label{AdSdynamiceq1}
(n-2)r\nabla_a\nabla_br=\frac{(n-2)(n-3)}{2}g_{ab}(1-\partial^c r\partial_c r) - r^2 T_{ab},
\end{equation}
\begin{equation}\label{AdSdynamiceq2}
T_{ab}=\partial_a\phi\partial_b\phi-\frac{1}{2}g_{ab}\partial^c\phi\partial_c\phi+g_{ab}\frac{(n-1)(n-2)}{2L^2},
\end{equation}
 where $L$ is the AdS radius. Observe that the system is no longer invariant under the following transformation
\begin{equation}\label{invarianttransformation}
r\mapsto cr, \qquad g_{ab}\mapsto c^2g_{ab}.
\end{equation}
Clearly, the last term in Eq. (\ref{AdSdynamiceq2}) does not respect the scaling symmetry.
However \cite{1994} shows that if the collapse of a spherecially symmetric scalar field admits homothetic symmetry, the system must be invariant under the transformation (\ref{invarianttransformation}).  Again we can easily extend the proof to arbitrary $n$ since it only concerns quantities on the quotient manifold. 
Therefore the homothetic symmetry is broken in $AdS_n$ for arbitrary $n$.

\subsection{Lightcone symmetry and quasi- self similarity in $AdS_n$}

Self similarity is explicitly broken in $AdS_n$. There are, however, certain evolutions in $AdS_n$ that admit other symmetries. Consider the following metric
\begin{equation}\label{AdSmetric}
ds^2=-2e^{2\sigma}dudv+\frac{r^2}{L^2}d\Omega^2,
\end{equation}
where $L$ is the $AdS$ radius and $\sigma$, $r$ are functions of $u,v$.

In assymtotically flat space the description of the system on quotient manifold essentially relies on the compatibility of the system, i.e. the angular components of Einstein equations are redundant, and the compatibility of the system is provided by twice contracted Bianchi identity \cite{christodoulou1986}, and since in $AdS$ the additional term in energy momentum tensor is proportional to the metric, the compatibility condition is still expected to hold.

Under the metric given in Eq. (\ref{AdSmetric}), the Einstein Klein Gordon system for general dimension $n$ is the following
\begin{equation}\label{AdSeq1}
\partial_u^2 r -2\partial_u\sigma\partial_u r +\frac{r}{n-2}(\partial_u\phi)^2=0,
\end{equation}
\begin{equation}\label{AdSeq2}
\partial_v^2 r -2\partial_v\sigma\partial_v r +\frac{r}{n-2}(\partial_v\phi)^2=0,
\end{equation}
\begin{equation}\label{AdSeq3}
r\partial_u\partial_v r = -\frac{n-3}{2}(e^{2\sigma}+2\partial_u r\partial_v r)+\frac{n-1}{2}e^{2\sigma}r^2,
\end{equation}
\begin{equation}\label{AdSeq4}
-2\partial_u\partial_v\sigma = \frac{1}{r}\partial_u\partial_v r-e^{2\sigma}(n-1)+\partial_u\phi\partial_v \phi,
\end{equation}
and KG equation
\begin{equation}\label{AdSeq5}
2r\partial_u\partial_v\phi+(n-2)(\partial_u r\partial_v \phi+ \partial_v r\partial_u \phi)=0.
\end{equation}

Observe that this system is invariant under the transformation
\begin{equation}\label{transformation}
u\mapsto au, \qquad v\mapsto a^{-1}v.
\end{equation}
Therefore it is natural to seek an isometry family and the corresponding generator of that family which can turn generically the set of PDE's into ODE's. In this case it is a second order dynamical system. To seek the corresponding solution, we assume the system only depends on the product of the two light-cone variables: $uv$. We denote such solutions as light-cone symmetric. A handful of solutions with such symmetries have been presented in \cite{Clement:2001ak}. This choice can give rise to quasi-self similar solutions by considering a change of variable, $v= -\frac{1}{k}$ and subsequently denoting $k$ by $v$. The metric becomes
\begin{equation}\label{quasi-similarmetric}
ds^2 = -\frac{2e^{2\rho}}{v^2}dudv+ r^2d\Omega^2.
\end{equation}
where $\rho$ and  $r$ are functions of $\frac{u}{k}$. Note in the Ansatz $\rho$ and $r$ only depend on $\frac{u}{v}$. Therefore it is a quasi self similar solution. Metric Eqs.(\ref{quasi-similarmetric}) is equvalent to $\alpha=0$ case in \cite{0264-9381-31-9-098001}, where $\alpha$ is a parameter in the introduced Ansatz. But only the $\alpha =0$ case of the quasi self similar solution is generalizable to higher dimensional $AdS$. 

However,  those solutions do not give rise to black holes without some proper coordinate extensions, since if we assume that the solution contains black hole, then on the apparent horizon, 
\begin{equation}
g^{uv}\partial_u r \partial_v r=\frac{2u}{v}e^{-2\rho}r'(\frac{u}{v})^2=0,
\end{equation}
but its sign can not change. Therefore, without some proper extension of coordinates, this form of the metric does not give rise to black hole solutions. In the \ref{App:constant scalar field} we exhibited the coordinate extension for static solutions. However the same extension does not apply to the dynamical case. Below we will only limit our discussion to the cases that the scalar field does not form a black hole. 

\subsection{The dynamical system in $AdS_n$}

Let $\eta=\frac{u}{v}$ and $\phi = \phi(\eta)-k\ln{\eta}$, where $k$ is a positive constant. The family $k=0$ and $k\neq 0$ are different solutions. The static cases, i.e. the scalar field is constant, can be solved analytically. For $n\geq 4$, it gives 

\begin{equation}\label{adsvaccummetric}
ds^2=-(1+r^2+\frac{2M}{r^{n-3}})dt^2+\frac{dr^2}{1+r^2+\frac{2M}{r^{n-3}}}+r^2d\Omega^2,
\end{equation}
where $M$ is a integration constant, and it agrees with $AdS$ vaccum metric in $M= 0$ and $AdS$ black hole when $M\neq 0$ as we expect.  For dynamical cases, we discuss both $k=0$ and $k\neq 0$ using dynamical system approaches. 

 \subsubsection{$k\neq0$}

Substitute the Ansatz $\eta=\frac{u}{v}$ and $\phi=\phi(\frac{u}{v})-k\ln{v}$ into Eqs.(\ref{AdSeq1}-\ref{AdSeq5}) and let $x=\ln{\eta}$, $y=e^{2\rho+x}$. The equations of motion take the following form:
\begin{equation}\label{dynamicalsystemAdSeq1}
\ddot{r}-\frac{\dot{y}}{y}\dot{r}+\frac{k^2r}{4(n-2)}=0,
\end{equation}
\begin{equation}\label{dynamicalsystemAdSeq2}
r\ddot{r}+(n-3)\dot{r}^2-\frac{y}{2}((n-1)r^2+n-3)=0.
\end{equation}

Let  $z=\sqrt{y}$ and $w=\frac{\dot{r}}{z}$. The dynamical system Eqs.(\ref{dynamicalsystemAdSeq1}-\ref{dynamicalsystemAdSeq2}) is given by
\begin{equation}\label{dynamicalsystemAdSc1eq1}
\dot{r}=wz,
\end{equation}
\begin{equation}\label{dynamicalsystemAdSc1eq2}
\dot{w}=\frac{z}{4r}((n-1)r^2+n-3)-\frac{(n-3)zw^2}{2r}-\frac{k^2r}{8(n-2)z},
\end{equation}
\begin{equation}\label{dynamicalsystemAdSc1eq3}
\dot{z}=\frac{z^2}{4rw}((n-1)r^2+n-3)-\frac{(n-3)z^2w}{2r}+\frac{k^2r}{8(n-2)w},
\end{equation}
 Note that $w=0$ corresponds to the apparent horizon. The positivity of the black hole mass requires that $w<1$ since in this case $w^2=g^{uv}\partial_u r \partial_v r$.  The system does not have fixed points and, therefore,  we turn to a numerical analysis below. Notice that, similar to the flat space case, the behavior of the system is generic in different dimensions, as the problem can be completely formulated on the quotient manifold. Therefore in the below numerical analysis we only take $n=5$ case. 

\begin{figure}[h]
\begin{center}
\includegraphics[width=2.5in,height=2.5in]{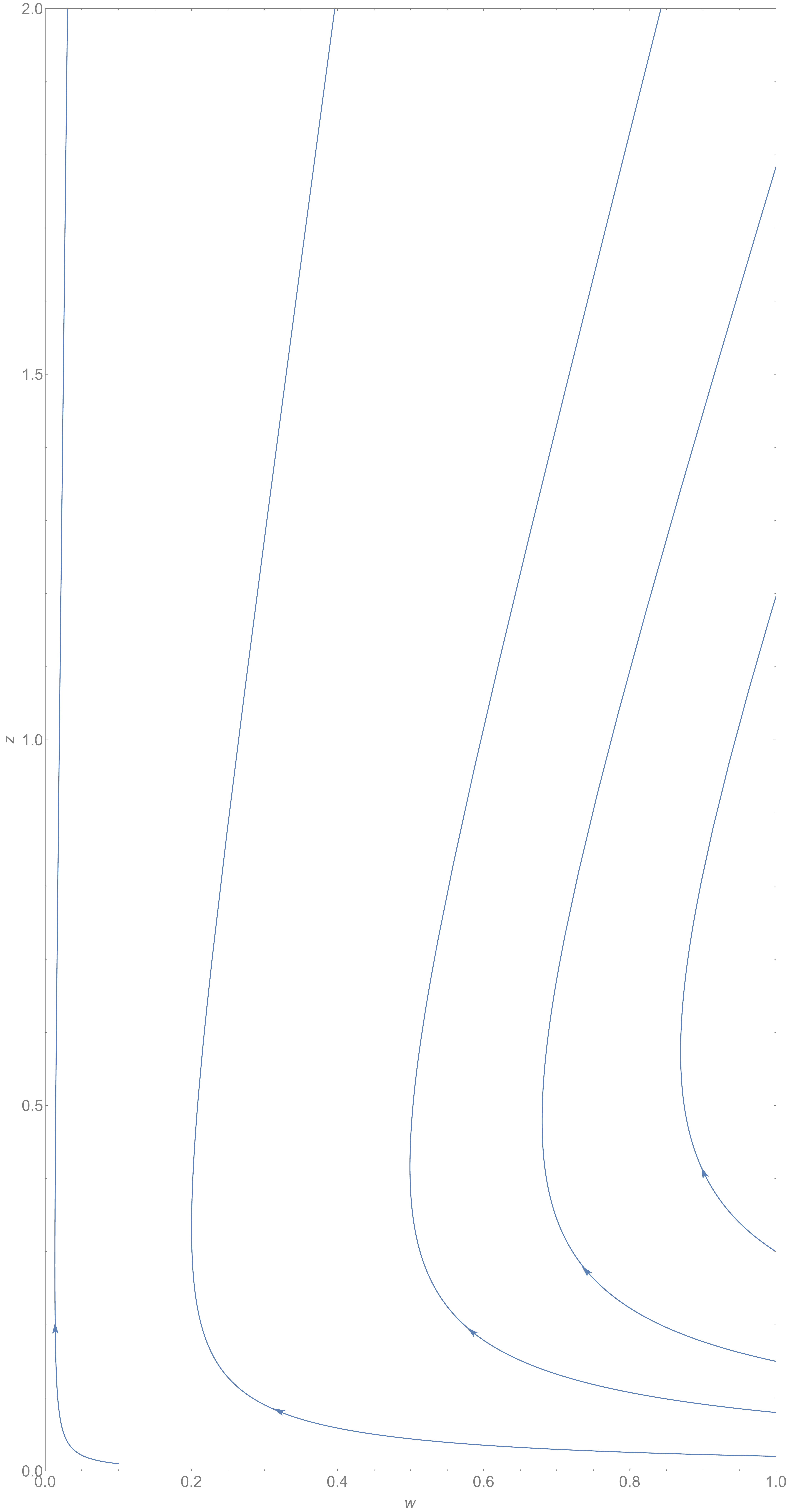}
\caption{\label{AdSphase1} Phase diagram for the case $k\neq 0$.For this diagram we take $n=5$ and $k=10$. We observe the scalar field evolves towards $w=0$ before bouncing to infinity.  }
\end{center}
\end{figure}

\begin{figure}[H]
\begin{center}
\includegraphics[width=2.5in,height=2.5in]{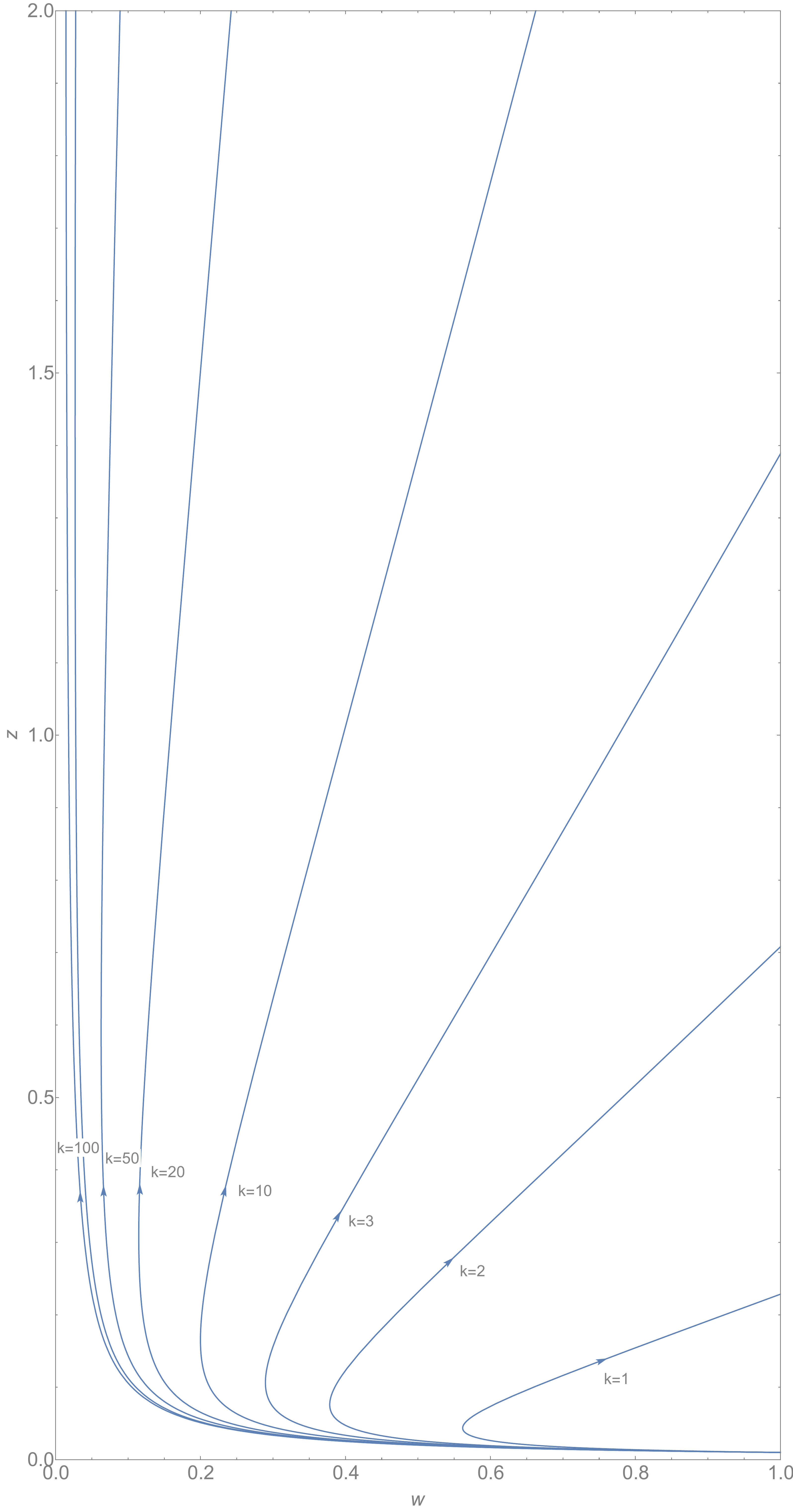}
\caption{\label{AdSphase2} The phase diagram for the influence of $k$ on the evolution of the scalar field. Qualitatively, the larger the $k$ is, the closer the evolution to form an apparent horizon, i.e. the smallest value of $w$ the evolution can attain is closer to $0$, where $w=0$ signifies an apparent horizon. When $k$ is large, the asymtotically behavior of the solution tends to be $w\to 0$ as $x\to \infty $. However, as Fig.(8) shows, there is no such change in asymtotical behavior of the system. We have , in all cases, $w\to \infty$ provided that $x$ is large enough. }
\end{center}
\end{figure}

\begin{figure}[H]
\begin{center}
\includegraphics[width=2.5in,height=2.5in]{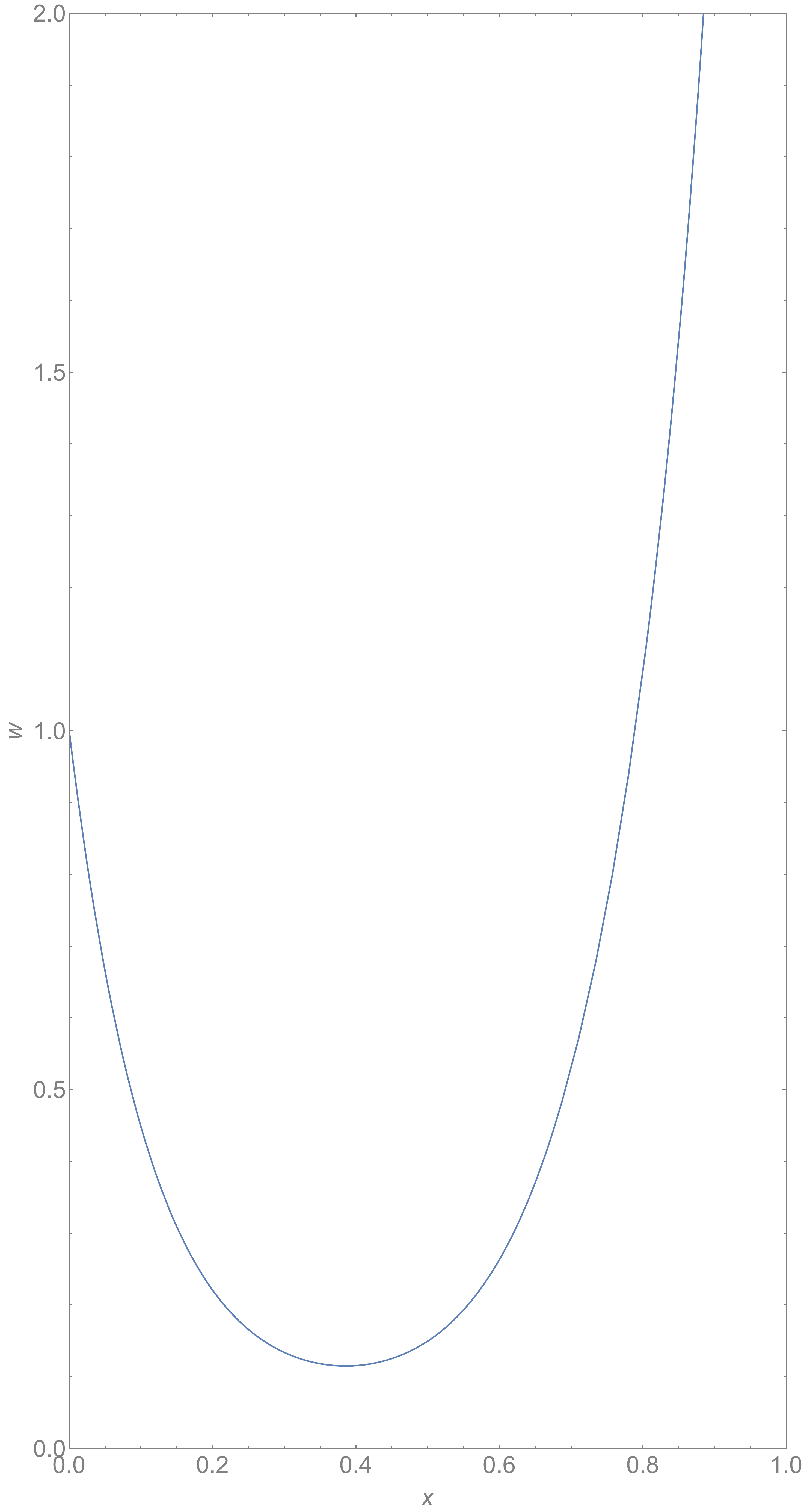}
\caption{\label{AdSphase3} For all $k$ we observe similar behavior of the solution $w$. Note $w$ decreases to a minimum then increases to infinty in a finite amount of time.   }
\end{center}
\end{figure}

Let $n=5$, the overall phase diagram is given in Fig.(\ref{AdSphase1}). As expected, there is no black hole solution. The parameter $k$ determines the strength of the scalar field. The phase diagram Fig.(\ref{AdSphase2}) shows the influence of $k$ on the phase evolution. It is evident that for $k$ large the evolution gets closer to the formation of an apparent horizon and subsequently black hole formation. However, Fig.(\ref{AdSphase3}) shows $w$ does not decrease to zero and therefore there is no apparent horizon in the system. Thus it is evident that no black hole is formed in the patch of quasi self similar solution we consider here. Notice that $w$ decreases to a minimum first, then increases to infinity in a finite amount of time. Thus the system should be interpted as scalar field interfering and then scattering to infinity in a finite amount of proper time. 

The reason we do not observe the formation of black hole, as discussed earlier, is that the specific quasi self-similar Ansatz we consider exclude the possibility of black hole formation if we do not extend our system in some coordinate extensions.

 \subsubsection{$k=0$}
Our treatment to the case of $k=0$ is similar to the one we used in last section. Substitute the Ansatz  $\eta=\frac{u}{v}$ and $\phi=\phi(\frac{u}{v})$ into Eqs.(\ref{AdSeq1}-\ref{AdSeq5}) and let $x=\ln{\eta}$, $y=e^{2\rho+x}$. The Einstein-Klein-Gordon equations are equivalently written as:

\begin{equation}\label{dynamicalsystemAdSeq1}
\ddot{r}-\frac{\dot{y}}{y}\dot{r}+\frac{A^2}{n-2}r^{5-2n}=0,
\end{equation}
\begin{equation}\label{dynamicalsystemAdSeq2}
r\ddot{r}+(n-3)\dot{r}^2-\frac{y}{2}((n-1)r^2+n-3)=0.
\end{equation}

Note the main difference comes from the exponent in $r$ in Eqs.(\ref{dynamicalsystemAdSeq1}).

Let  $z=\sqrt{y}$ and $w=\frac{\dot{r}}{z}$. The dynamical system is given by

\begin{equation}\label{dynamicalsystemAdSc2eq1}
\dot{r}=wz,
\end{equation}
\begin{equation}\label{dynamicalsystemAdSc2eq2}
\dot{w}=\frac{z}{4r}((n-1)r^2+n-3)-\frac{(n-3)zw^2}{2r}-\frac{A^2r^{5-2n}}{2(n-2)z},
\end{equation}
\begin{equation}\label{dynamicalsystemAdSc2eq3}
\dot{z}=\frac{z^2}{4rw}((n-1)r^2+n-3)-\frac{(n-3)z^2w}{2r}+\frac{A^2r^{5-2n}}{2(n-2)w}.
\end{equation}

We made the phase diagram of this system when $n=5$ and it is similar to (\ref{AdSphase1}), and the behavior of solution to the system for $k=0$ is similar to those to $k\neq 0$.

\begin{figure}[htp]
\begin{center}
\includegraphics[width=2.5in,height=2.5in]{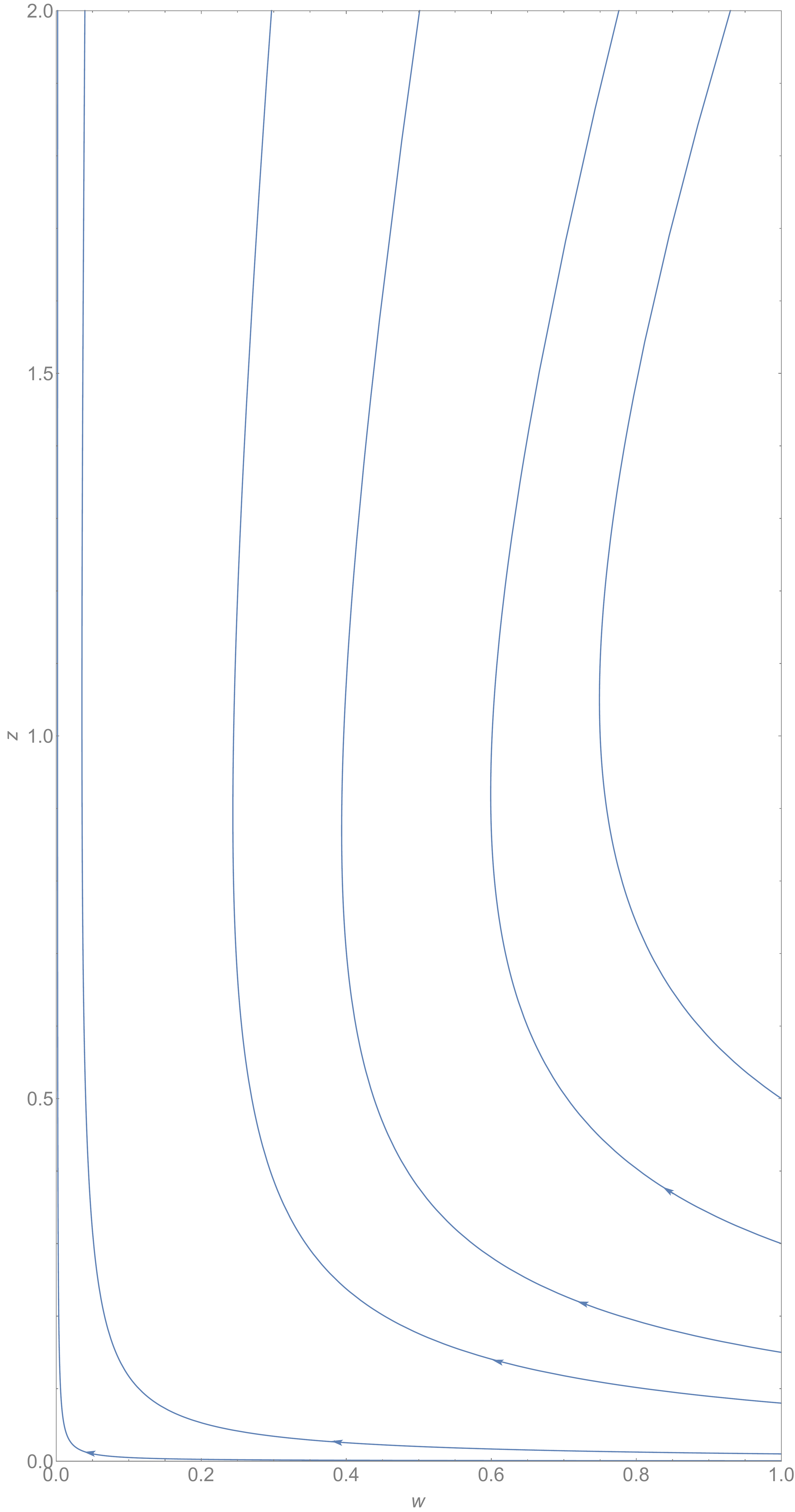}
\caption{\label{AdSphase4} Phase diagram for the case $k=0$.Note the phase diagram is similar to that of $k\neq 0 $ case.   }
\end{center}
\end{figure}

\begin{figure}[htp]
\begin{center}
\includegraphics[width=2.5in,height=2.5in]{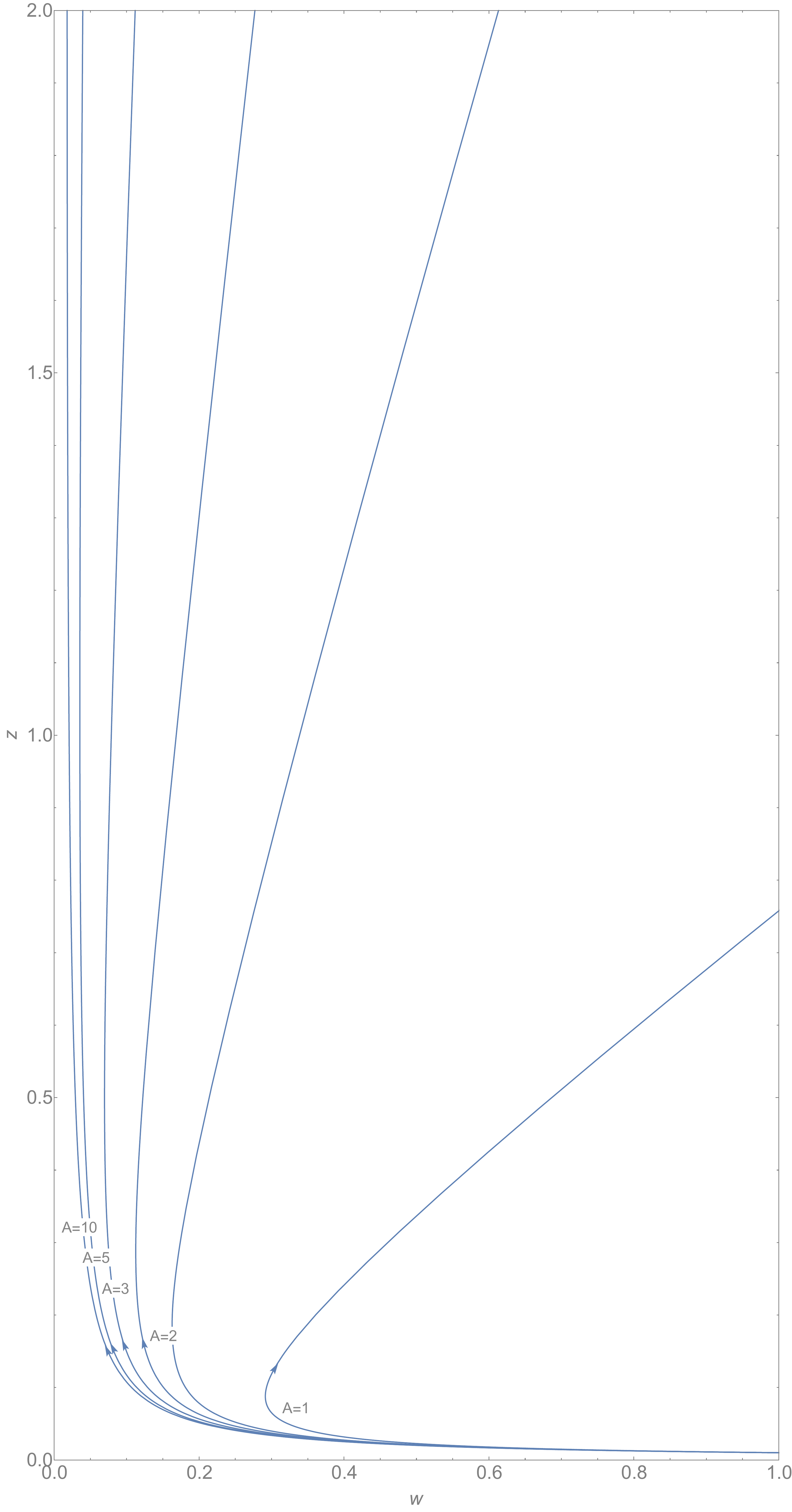}
\caption{\label{AdSphase5} The phase diagram for the influence of $k$ on the evolution of the scalar field for the case $k=0$. We observe similar behavior as those in the case $k\neq 0$. Note the scalar field collapses and then bounces back to infinity in a finite amount of time. Notice that the larger $A$ is, the closer the evolution is to reach $w=0$ which signifies the formation of an apparent horizon. However as Fig.(11) shows, there's no solution whose asymtotical behavior is $w\to 0$ as $x\to \infty$.   }
\end{center}
\end{figure}

\begin{figure}[htp]
\begin{center}
\includegraphics[width=2.5in,height=2.5in]{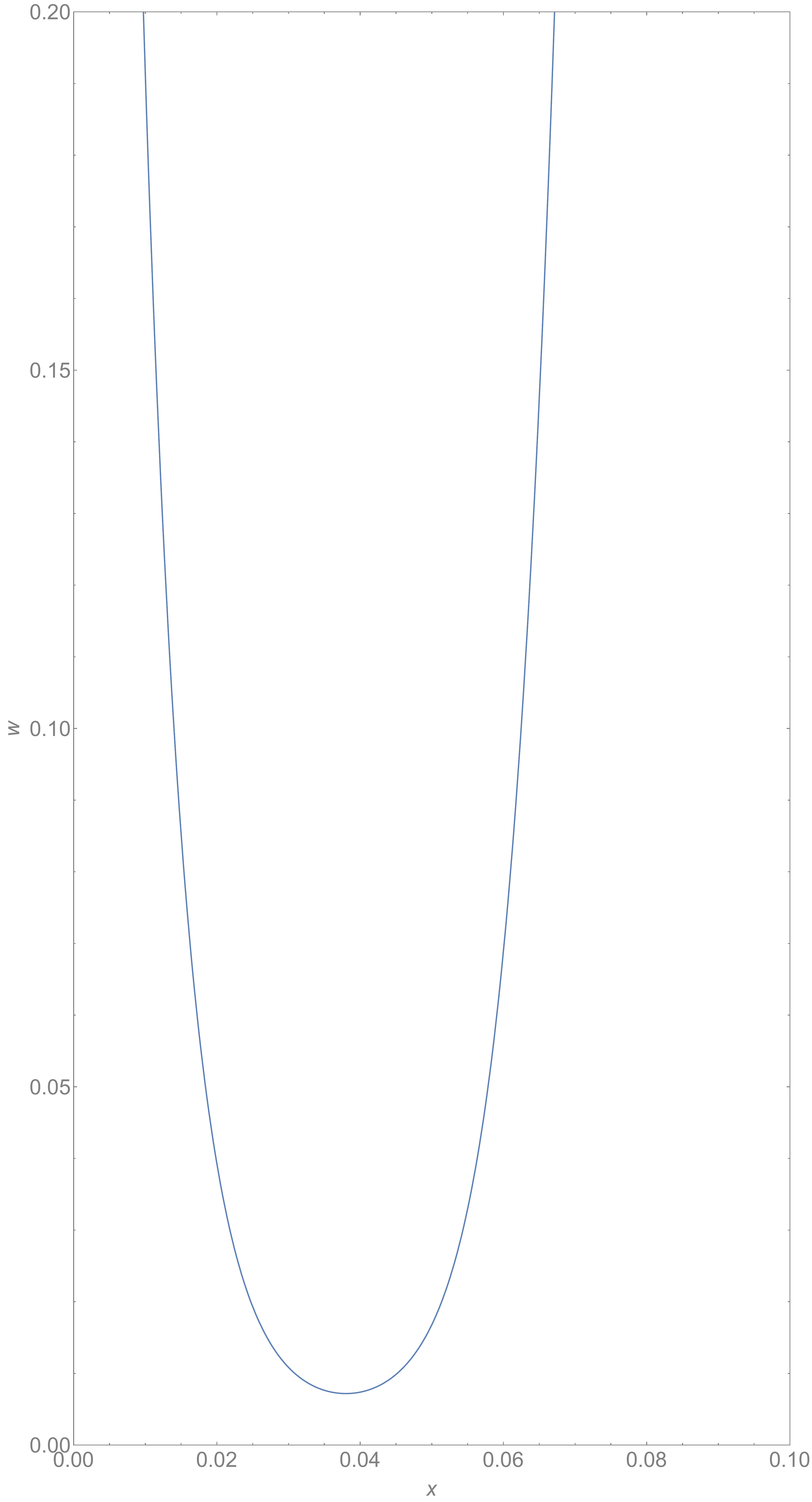}
\caption{\label{AdSphase4} Notice that for the case $k=0$, $w$ decreases to a minimum first and then increases to infinity in a finite amount of time. We observe that $w$ does not reach $0$, i.e. no apparent horizon is forned.   }
\end{center}
\end{figure}

Notice that the asymtotic behavior of the $k=0$ and $k\neq 0$ dynamical system should be different as we can observe from the difference in the last terms of Eq.(\ref{dynamicalsystemAdSeq1}) and Eq.(\ref{dynamicalsystemAdSeq1}). However, as the numerical evidence shows, they differ in a way such that the last terms of Eq.(\ref{dynamicalsystemAdSeq1}) and Eq.(\ref{dynamicalsystemAdSeq1}) tend to zero as $t\to \infty$, i.e. they approach to $AdS$ vaccum. This observation and the lack of apparent horizons in evolutions confirm our intrepertation of the solutions as interfering and scattering of the scalar field to the infinity in a finite amount of proper time. Thus although we are not able to observe black hole formation, a problem rooted in our forms of Ansatz, we have shown that the quasi self-similar behavior of these solutions does allow for a semi-analytic understanding.

\section{Conclusions}\label{Sec:Conclusions}

In this paper we have studied self-similarity and quasi self similarity of the Einstein-Klein-Gordon system in asymptotically flat and asymptotically AdS spacetimes.  In the case of asymptotically flat spacetimes we extended the previous result of \cite{Soda1996271} to self similar solution of type $k$ and observed a new transition from fixed points to black holes. Applying the standard techniques we obtained, to our knowledge, a new critical exponent in the higher dimensions when $k\neq0$. 

In the case of asymptotically AdS spacetimes there are not self simialar solutions. Nevertheless, inspired by the treatment of quasi self similar solutions in \cite{Clement:2001ak} and \cite{Baier:2014}, we generalized the quasi self similar solution into higher dimensional $AdS$. But different from the analytical methods used extensively in $AdS_3$, we turned the system into a dynamical system similar to that in asymtotically flat space and apply standard dynamical system techiques. The analysis provided a semi-analytical understanding of the interations of the scalar field that scatters to infinity in a finite amount of time. 

Yet it should be remarked that our dynamical system does not contain black hole solutions. But even in the static case, as the appendix shows, there could be black hole solutions if one extends the coordinate properly. Therefore the fact that the above dynamical system does not contain black holes may be due to a bad choice of coordinate preventing black hole formations. The idea of quasi self similarity is heavily coordinate dependent. For example, one may consider the following Bondi coordinate, which in flat space provides a complete charaterization of the collapse of the self similar scalar field, 

\begin{equation}
ds^2= -\bar{g}gdu^2-2gdudr+r^2d\Omega^2
\end{equation}

But the lightcone symmetry and the quasi self similarity manifestly lost for $n>3$. Therefore a complete charaterization of the quasi self similar collapse of a spherically symmetric scalar field still remains an open problem.

\section*{Acknowledgments}
I am particularly thankful to Leo Pando Zayas for suggesting this problem and guidance throughout the course of this project. 
I am also thankful to David Garfinkle for various comments. This work is  partially supported by Department of Energy under grant DE-SC0007859.

\appendix
\section{The derivation of the dynamic equations for thel collapse of a self gravitating massless scalar field}
Write (\ref{Bondimetric}) in terms of non coordinate bases,
\begin{equation}\label{noncoordinatebase}
ds^2=\eta_{ab}e^a\otimes e^b
\end{equation}
Let$e^0=\frac{g+\bar{g}}{2}du+dr$ and $e^1=\frac{\bar{g}-g}{2}du+dr$, where $g$ and $\bar{g}$ are functions of u and r.
Let $e^i=r\sin\theta_1\sin\theta_2\dots\sin\theta_{i-2}d\theta_{i-1}$, where $i$ ranges from 2 to $n-1$. Moreover, we will write indices larger than 1 with $i$ and $j$.

Cartan's structure equations imply the spin connections as follows
\begin{equation}\label{connection1}
\omega^0_1=\omega^1_0=\frac{1}{2}(\partial_r\bar{g}+\frac{\bar{g}}{g}\partial_r g)du+\frac{1}{g}\partial_r g dr
\end{equation}
\begin{equation}\label{connection2}
\omega^i_0=\omega^0_i=\frac{g-\bar{g}}{2g}\sin\theta^1\sin\theta^2\dots\sin\theta^{i-2}d\theta^{i-1}
\end{equation}
\begin{equation}\label{connection3}
\omega^i_1=-\omega^1_i=\frac{\bar{g}+g}{2g}\sin\theta^1\dots\sin\theta^{i-2}d\theta^{i-1}
\end{equation}
\begin{equation}\label{connection4}
\omega^i_j=\cos\theta^{j-1}\sin\theta^{j}\dots\sin\theta^{i-2}d\theta^{i-1}
\end{equation}

For the curvature 2 form,  $R^k_k=0$ $\forall$ $k$.We will use alphabetical letter to denote non-coordinate frame indices and Greek letters coordinate indices.
Observe that there are only 4 independent components of curvature 2 form that are not associated by some functions of angles with each other. Therefore a direct computation gives
\begin{equation}\label{curvature2form1}
R^0_1=(\partial_u(\frac{1}{g}\partial_r g)-\frac{1}{2}(\partial^2_r\bar{g}+\partial_r(\frac{\bar{g}}{g}\partial_r g)))du\wedge dr
\end{equation}
\begin{equation}\label{curvature2form2}
R^0_i=(\partial_u \frac{g-\bar{g}}{2g}-\frac{g+\bar{g}}{4g}(\partial_r\bar{g}+\frac{\bar{g}}{g}\partial_r g))\sin\theta_1\dots\sin\theta_{i-2}du\wedge d\theta^{i-1}
\end{equation}
\[+(\partial_r \frac{g-\bar{g}}{2g}-\frac{g+\bar{g}}{2g^2}\partial_r g)\sin\theta_1\dots\sin\theta_{i-2}dr\wedge d\theta^{i-1}\]
\begin{equation}\label{curvature2form3}
R^1_i=(\frac{g-\bar{g}}{4g}(\partial_r \bar{g}+\frac{\bar{g}}{g}\partial_r g)-\partial_u \frac{\bar{g}+g}{2g})\sin\theta_1\dots\sin\theta_{i-2}du\wedge d\theta^{i-1}
\end{equation}
\[
+(\frac{g-\bar{g}}{2g^2}\partial_r g-\partial_r\frac{\bar{g}+g}{2g})\sin\theta_1\dots\sin\theta_{i-2}dr\wedge d\theta^{i-1}
\]
\begin{equation}\label{curvature2form4}
R^i_j=(1-\frac{\bar{g}}{g})\sin^2\theta_1\dots\sin^2\theta_{j-2}\sin\theta_{j-1}\dots\sin\theta_{i-2}d\theta^{i-1}\wedge d\theta^{j-1}
\end{equation}
where in the above $i>j$
Note that it can be shown those components of curvature 2 form satisfy the Bianchi identity.Compute $R^a{}_{bcd}$ in co frame. They are given by
\begin{equation}\label{riemann1}
R^0{}_{101}=\frac{1}{2g^3}(-\bar{g}(\partial_r g)^2+g\partial_r g\partial_r \bar{g}+g\bar{g}\partial^2_rg+g^2\partial^2_r\bar{g}+2\partial_r g\partial_u g-2g\partial_r\partial_u g)
\end{equation}
\begin{equation}\label{riemann2}
R^0{}_{i0i}=-(\frac{\partial_r g}{4rg}+\frac{\bar{g}^2}{4rg^3}\partial_r g+\frac{\partial_r \bar{g}}{2rg}-\frac{\bar{g}\partial_u g}{2rg^3}+\frac{\partial_u \bar{g}}{2rg^2})
\end{equation}
\begin{equation}\label{riemann3}
R^1{}_{i1i}=-R^0{}_{i0i}-\frac{\partial_r \bar{g}}{rg}
\end{equation}
\begin{equation}\label{riemann4}
R^0{}_{i1i}=-(\frac{\partial_r g}{4rg}-\frac{\bar{g}^2\partial_r g}{4rg^3}+\frac{\bar{g}\partial_u g}{2rg^3}-\frac{\partial_u \bar{g}}{2rg^2})
\end{equation}
\begin{equation}\label{riemann5}
R^1{}_{i0i}=-R^0{}_{i1i}
\end{equation}
\begin{equation}\label{riemann6}
R^i{}_{jij}=\frac{1}{r^2}(1-\frac{\bar{g}}{g})
\end{equation}
It is a direct computation to get Einstein tensor out of (\ref{riemann1}-\ref{riemann6}) and transform it into the Bondi coordinate. Therefore the Einstein tensors are given as follows
\begin{equation}\label{Einstein1}
G_{00}=(n-2)(\bar{g}^2\frac{\partial_r g}{2rg}-\frac{\bar{g}\partial_u g}{2rg}+\frac{\partial_u \bar{g}}{2r}-\frac{\bar{g}\partial_r\bar{g}}{2r}+\frac{n-3}{2r^2}(g-\bar{g})\bar{g})
\end{equation}
\begin{equation}\label{Einstein2}
G_{01}=(n-2)(\bar{g}\frac{\partial_r g}{2rg}-\frac{\partial_r \bar{g}}{2r}+\frac{(n-3)(g-\bar{g})}{2r^2}
\end{equation}
\begin{equation}\label{Einstein3}
G_{11}=(n-2)\frac{\partial_r g}{rg}
\end{equation}

\section{The stability and the corresponding metric of the fixed points}\label{App:Stability}
For the case $u<0$, the fixed point for the system (\ref{dynamiceq1}-\ref{dynamiceq3}) can be found by directly solving the corresponding algebraic equations. Note that there is another fixed point for the system $z=0, y=n-3, \gamma=0$ but spacetime at that point is singular. Therefore the fixed points of interests are only $(\frac{\sqrt{n-2}}{k+\sqrt{n-2}},\frac{n-3}{n-2},\sqrt{n-2})$ and $ (-\frac{\sqrt{n-2}}{k-\sqrt{n-2}},\frac{n-3}{n-2},-\sqrt{n-2})$ in terms of $(z,y,\gamma)$. And the following will show they are of similar nature therefore it is enough to discuss one of them.

Denote z coordinate of two fixed points as $z_1$ and $z_2$.The characteristic polynomial for both is
\begin{equation}\label{characteristicpolynomial}
\lambda^2+\frac{(n-2)(1-z_{1,2})}{1-2z_{1,2}}\lambda+\frac{2(n-2)}{1-2z_{1,2}}
\end{equation}
Therefore the eigen values are given by
\begin{equation}\label{eigenvalue}
\lambda_{1,2}=\frac{(n-2)(1-z_{1,2})}{2(2z_{1,2}-1)}\pm\sqrt{\frac{(n-2)^2(1-2z_{1,2})+8(n-2)(2z_{1,2}-1)}{4(1-2z_{1,2})^2}}
\end{equation}
Thus there are two classes depending on the value of $z_{1,2}$ which is determined by $k$. When $k^2\geq n-2$, $\lambda_{1,2}$ are both negative and when $k^2<n-2$, $\lambda_{1,2}$ are both positive.

Notice that fixed points represent conformally flat spacetime as trivial solutions to the Einstein equation. Without losing of generality, we focus our discussion on $(\frac{\sqrt{n-2}}{k+\sqrt{n-2}},\frac{n-3}{n-2},\sqrt{n-2})$. Substitute it back we get
\begin{equation}\label{App:Fixedpointspacetime}
\bar{g}=\frac{k+\sqrt{n-2}}{\sqrt{n-2}}\frac{r}{u}, g=\frac{\sqrt{n-2}}{n-3}(k+\sqrt{n-2})\frac{r}{u}
\end{equation}
The corresponding metric is given by
\begin{equation}\label{App:Fixedpointmetric}
ds^2=-\frac{(k+\sqrt{n-2})^2}{n-3}\frac{r^2}{u^2}du^2-\frac{2\sqrt{n-2}}{n-3}(k+\sqrt{n-2})\frac{r}{u}dudr+r^2d\Omega^2
\end{equation}
Let $t=\frac{k+\sqrt{n-2}}{\sqrt{n-3}}\ln{u}+x$, $x=\sqrt{n-2}{n-3}\ln{r}$. The metric (\ref{App:Fixedpointmetric}) is turned to
\begin{equation}\label{App:ConformalMetric}
ds^2=e^{\sqrt{\frac{n-3}{n-2}}x}(-dt^2+dx^2+d\Omega^2)
\end{equation}
\section{The case for constant scalar field in $n\geq 4$}\label{App:constant scalar field}
Let $k=0$ or $A=0$ in Eqs. \ref{dynamicalsystemAdSeq1}. Then it is readily integrable and it gives
\begin{equation}
y=M\dot{r}
\end{equation}
The equation (\ref{dynamicalsystemAdSeq2}) can be integrated as follows
\begin{equation}\label{constantscalarfield}
r^{n-3}\dot{r}-\frac{M}{2}(r^{n-1}+r^{n-3})=N
\end{equation}
Directly solve Eqs.(\ref{constantscalarfield}), we get
\begin{equation}\label{Adsvaccumint}
x= \int \frac{dr}{\frac{M}{2}(1+r^2)+\frac{k}{r^{n-3}}}+C=F(r)
\end{equation}
The integral (\ref{Adsvaccumint}) is in general not able to be expressed into a closed form. We may perform the following coordinate transformation to avoid evaluating the integral
\begin{equation}\label{coordinatetransformation}
u=e^{\frac{1}{2}(t+F(r))}, v=e^{\frac{1}{2}(t-F(R))}
\end{equation}
Substitute the transformation (\ref{coordinatetransformation}) into the metric, one gets AdS vaccum metric:
\begin{equation}\label{adsvaccum}
ds^2=-(1+r^2+\frac{2M}{r^{n-3}})dt^2+\frac{dr^2}{1+r^2+\frac{2M}{r^{n-3}}}+r^2d\Omega^2
\end{equation}
where we rescale $t$ and absorb one of the integral constant.
\section{Perturbation around $AdS_{n}$ vaccum solution}
We may consider perturbing the $AdS_{n}$ vaccum solution to obstain the spacetime corresponding the begining of the evolution, i.e. when the scalar field strength is small. Consider let $y=e^z$. Let $z=z_0+\delta z$, where $z_0$ is the vacuum solution and $\delta z$ is small. Thus to the lowest order $e^z\sim e^{z_0}$. Rewrite the system as follows
For $k\neq0$
\begin{equation}\label{dynamicalsystemAdSeq1app}
\ddot{r}-\dot{z_0+\delta z}\dot{r}+\frac{k^2r}{4(n-2)}=0,
\end{equation}
\begin{equation}\label{dynamicalsystemAdSeq2app}
r\ddot{r}+(n-3)\dot{r}^2-\frac{e^{z_{0}}}{2}((n-1)r^2+n-3)=0.
\end{equation}
Thus we solve
\begin{equation}
\delta y=\int \frac{rk^2dr}{4(n-2)\dot{r}^2}=\int \frac{rk^2 dr}{4(n-2)(1+r^2+\frac{M}{r^n-3})^2}.
\end{equation}
Thus we can write the metric as
\begin{equation}
ds^2=-(1+r^2+\frac{M}{r^{n-3}}+\frac{k^2}{4(n-2)}G(r))dt^2+\frac{(1+r^2+\frac{M}{r^{n-3}}+\frac{k^2}{4(n-2)}G(r)dr^2}{(1+ r^2+\frac{M}{r^{n-3}})^2}+r^2d\Omega^2.
\end{equation}

where $G(r)=\int \frac{rdr}{(1+ r^2+\frac{M}{r^{n-3}})^2}$.

and for $k=0$
\begin{equation}\label{dynamicalsystemAdSeq1app}
\ddot{r}-\dot{z_{0}+\delta z}\dot{r}+\frac{A^2}{n-2}r^{5-2n}=0,
\end{equation}
\begin{equation}\label{dynamicalsystemAdSeq2app}
r\ddot{r}+(n-3)\dot{r}^2-\frac{e^{z_0}}{2}((n-1)r^2+n-3)=0.
\end{equation}
A similar approximation shows that the metric is approximated by
\begin{equation}
ds^2=-(1+r^2+\frac{M}{r^{n-3}}+{\frac{A^2}{(n-2)}G(r)})dt^2+\frac{(1+M_1r^2+\frac{M_2}{r^{n-3}}+\frac{A^2}{(n-2)}G(r))dr^2}{(1+M_1 r^2+\frac{M_2}{r^{n-3}})^2}+r^2d\Omega^2.
\end{equation}

where $G(r)=\int \frac{dr}{r^{2n-5}(1+M_1r^2+\frac{M_2}{r^{n-3}})^2}$.

\bibliographystyle{JHEP}
\bibliography{paper_new,Collapsebib}

\end{document}